\documentclass[10pt,letterpaper]{article}

\usepackage{ccn}
\usepackage{hyperref}
\usepackage{pslatex}
\usepackage{apacite}
\usepackage{booktabs}
\usepackage{graphicx, multicol, latexsym, amsmath, amssymb}
\usepackage{caption}
\usepackage{subcaption}
\usepackage{dblfloatfix}

\title{Learning Intermediate Features of Object Affordances \\ with a Convolutional Neural Network}
 
\author{{\large \bf Aria Y. Wang (yuanw3@andrew.cmu.edu)} \\
  Center for Neural Basis of Cognition (CNBC), Carnegie Mellon University\\
5000 Forbes Ave, Pittsburgh, PA 15213 United States
  \AND {\large \bf Michael J. Tarr (michaeltarr@cmu.edu)} \\
  Department of Psychology, Carnegie Mellon University\\
5000 Forbes Ave, Pittsburgh, PA 15213 United States}

\begin{document}

\maketitle

\section{Abstract}
{
\bf
Our ability to interact with the world around us relies on being able to infer what actions objects afford -- often referred to as affordances. The neural mechanisms of object-action associations are realized in the visuomotor pathway where information about both visual properties and actions is integrated into common representations. However, explicating these mechanisms is particularly challenging in the case of affordances because there is hardly any one-to-one mapping between visual features and inferred actions. To better understand the nature of affordances, we trained a deep convolutional neural network (CNN) to recognize affordances from images and to learn the underlying features or the dimensionality of affordances. Such features form an underlying compositional structure for the general representation of affordances which can then be tested against human neural data. We view this representational analysis as the first step towards a more formal account of how humans perceive and interact with the environment. }
 
\begin{quote}
\small
\textbf{Keywords:} 
affordance; dataset; convolutional neural network;
\end{quote}

While interacting with our environment, we naturally infer the functional properties of the objects around us. These properties, typically referred to as affordances, are defined by~\citeA{gibson1979theory}, as all of the actions that an object in the environment offers to an observer. For example, “kick” for a ball and “drink” for water. Understanding affordances is critical for understanding how humans are able to interact with objects in the world.

In recent years, convolutional neural networks have been successful in preforming object recognition in large-scale image datasets~\cite{krizhevsky2012imagenet}. At the same time, convolutional networks trained to recognize objects have been used as feature extractors and can successfully model neural responses as measured by fMRI in human visual cortex~\cite{agrawal2014pixels} or by electrodes in monkey IT cortex~\cite{yamins2016using}. To understand the relevant visual features in an object that are indicative of affordances, we trained a CNN to recognize affordable actions of objects in images.

\section{Dataset Collection}
\begin{figure}[!h]
    \centering
    \includegraphics[width=0.4\textwidth]{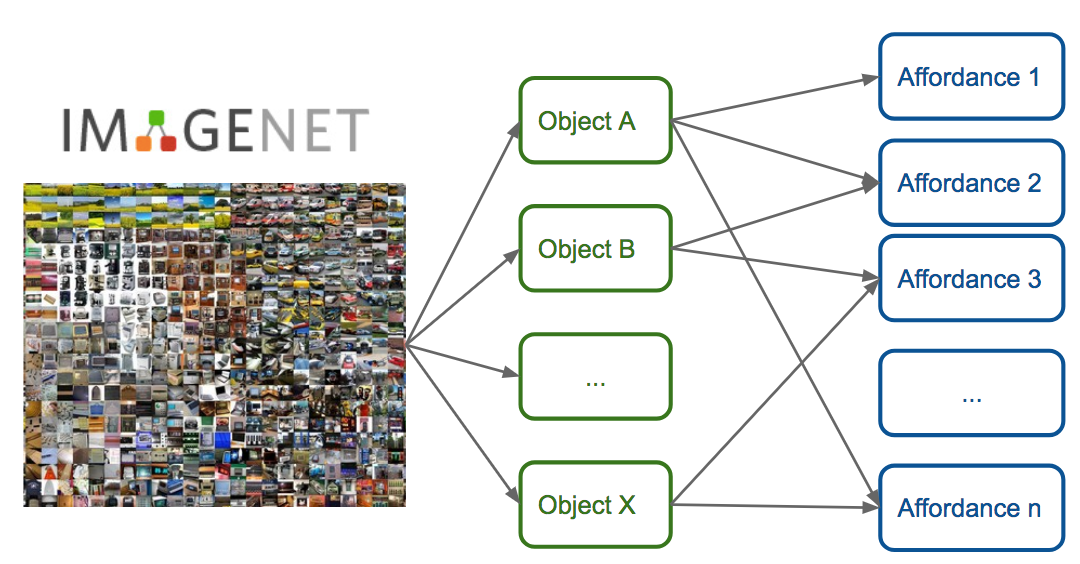}
    \caption{Dataset Collection. The labelers are given object labels, indicated in the green boxes here and assign to them affordances labels, indicated with blue boxes.}
    \label{fig:data}
\end{figure}

Training deep CNNs is known to require large amounts of data. Available affordance datasets with images and semantic labels are largely limited at this moment. The only relevant dataset currently available to the public was created by~\citeA{chao2015mining}, and only includes affordance labels for 20 objects from the PASCAL dataset and 90 objects from the COCO dataset. Here we built a large scale affordance dataset with affordances labels attached to all images in the ImageNet dataset~\cite{deng2009imagenet}. This dataset forms a more general representation of the affordance space and allows large scale end-to-end training from the image space and to this affordance space. The dataset collection process is shown in Figure~\ref{fig:data}. Human labelers were presented with object labels from ImageNet object categories and answered the question ``What can you do with that object?''. All answers were then co-registered with WordNet~\cite{miller1995wordnet} action labels so that our labels could be extended to other datasets. The top five  responses from labelers were used as canonical affordance labels for each object. 334 categories of actions were labeled for around 500 objects categories. When combined with image to object label mappings from ImageNet, these affordance labels provided us with the image to affordance label mappings that were used to train our CNN. 

\section{Visualization of Affordance Space}
\begin{figure}[!h]
    \centering
    \includegraphics[width=0.55\textwidth]{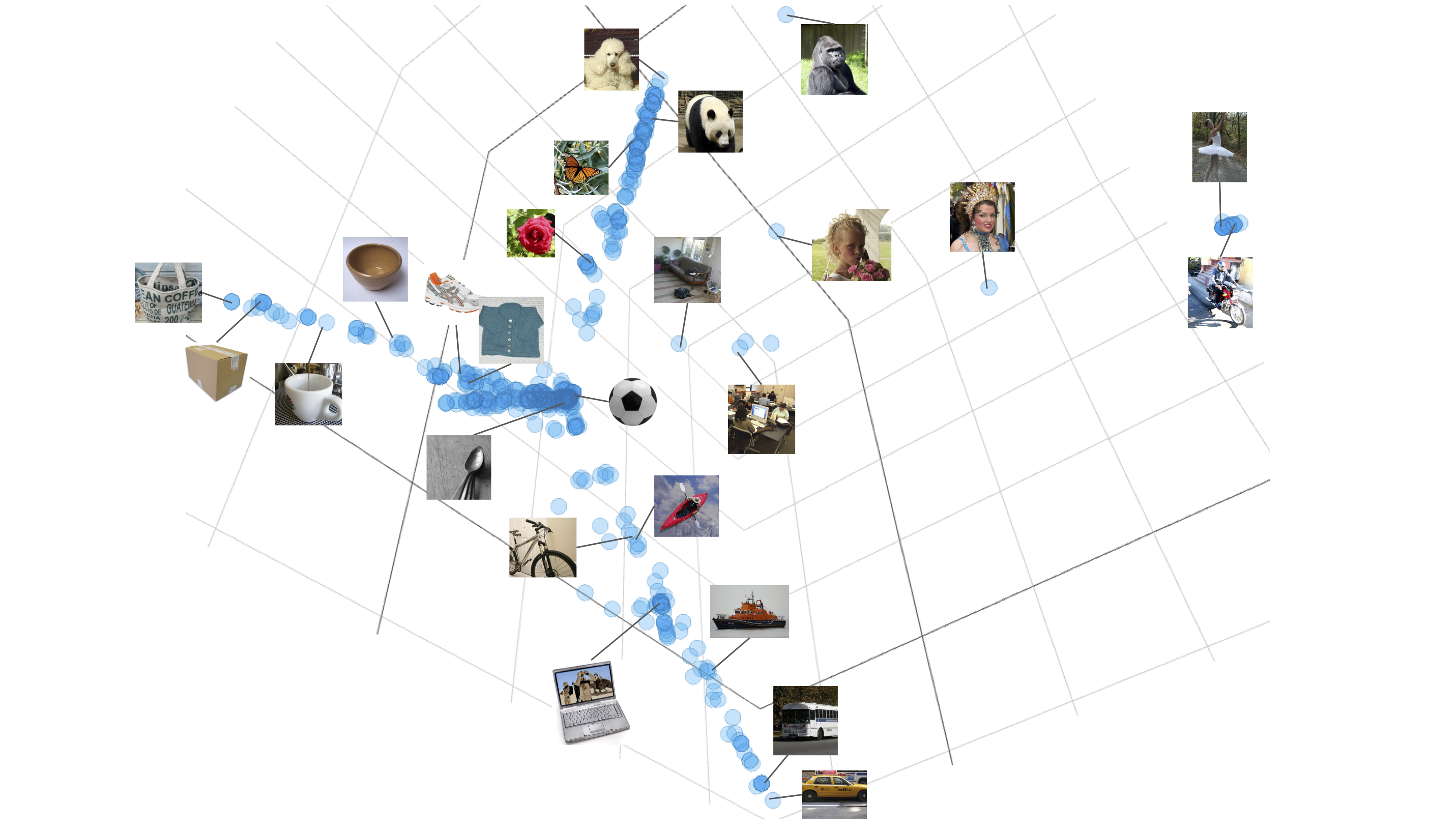}
    \caption{ImageNet images in the affordance space.}
    \label{fig:aff_space}
\end{figure}

In our affordance dataset, each object was represented by a binary vector indicating whether each of the possible actions was available for this object or not. Each object can then be represented as a point in the affordance space. We used PCA to project these affordance vectors into a 3D space and plotted the object classes as illustrated in Figure~\ref{fig:aff_space}. In the 3D space created for visualization, the objects appear to be well separated. More specifically, the majority of living things were organized along the top axis; the majority of small household items were organized along the left axis; and transportation tools and machines were organized along the right axis. Human-related categories such as dancer and queen do not belong to any axis and appear as flowing points in the space.
\begin{table*}[!b]
\begin{center} 
\caption{Training Results. ``Fine-tuning'' indicates that the network was pre-trained to predict image categories, while ``Training from Scratch'' indicates that the network was initialized with random weights. Baseline accuracy was calculated by estimating the most frequent categories.} 
\label{result-table} 
\vskip 0.12in
\begin{tabular}{@{}llllll@{}} 
\toprule
& Baseline &  Fine-tuning & Training from  & Fine-tuning  & Training from  \vspace{-0.5mm} \\
&&& Scratch & w/ oversampling & scratch w/ oversampling
 \\
\midrule
Training Accuracy (\%) & 7.61 & 80.39 & 71.42 & 87.60 & 85.05\\
Testing Accuracy (\%) & 6.86 & 44.62 & 37.47 & \textbf{55.42} & \textbf{53.43} \\
\bottomrule
\end{tabular} 
\end{center} 
\end{table*}

\section{Results}
\subsection{Network Training}
A CNN was trained to predict affordance categories from images. A total of 55 affordances were selected as potential actions after ensuring that each affordance label had at least 8 object categories associated with it (by removing affordances that were associated with too few object categories). Each object category was placed in the training, validation or testing sets. These sets were exclusive, such that, if one object category appeared in one set, it would not appear in the other two sets. Such separation ensures that the learning of affordances was not based on recognizing the same objects and learning linear mappings between objects and affordances.

We used the ResNet18 model~\cite{he2016deep} (other models such as VGG produced similar results), and trained it using the Adam optimizer~\cite{kingma2014adam} by minimizing binary cross-entropy loss. Approximately 630,000 images from ImageNet were used in training, and approximately 71,000 images each were used for validation and testing. The trained CNN was evaluated by computing the average percentage of correctly predicted affordance labels, and the results are reported in Table~\ref{result-table}. The trained networks showed significantly better performance compared to the baseline.

\subsection{Skewed Distribution and Oversampling}
\begin{figure}[!h]
    \centering
    \includegraphics[width=0.45\textwidth]{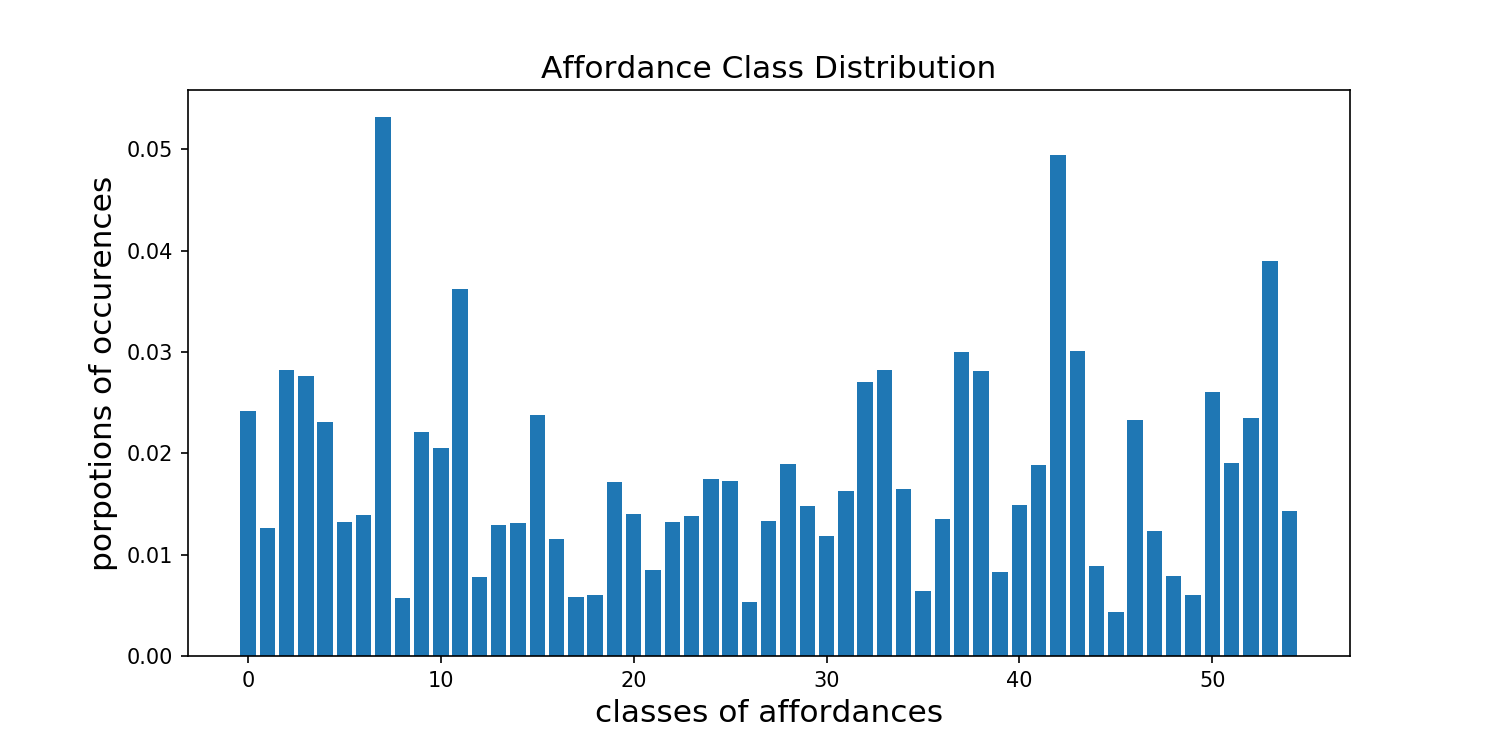}
    \caption{Percentages of objects classes assigned to each affordances categories.}
    \label{fig:dist}
\end{figure}

Since actions such as ``hold'' and ``grab'' would be used on objects much more often than actions such as ``thrust'', we obtained an uneven distribution of affordance labels across image categories, as shown in Figure~\ref{fig:dist}. In computer vision, oversampling is a commonly used solution for this problem. However, because of the multi-label nature of the affordance recognition problem, proper oversampling is challenging. Less frequently appearing classes need to be oversampled without over representing the more frequently appearing classes. We used Multi-label Best First Over-sampling (ML-BFO)~\cite{ai2015best}, and re-trained the CNN with the resampled data. This produced a considerable increase in prediction performance, as seen in Table~\ref{result-table}.

\subsection{Sample Predictions}
\begin{figure}[!h]
    \centering
    \begin{subfigure}[b]{0.14\textwidth}
        \centering
        \includegraphics[width=\textwidth]{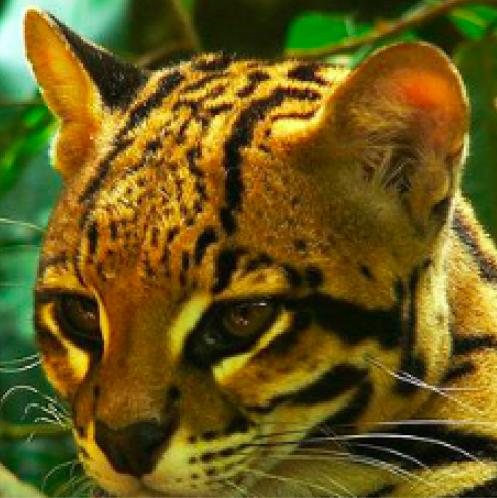}
        \caption{care/feed}
    \end{subfigure}
    \hspace{0.3cm}
    \begin{subfigure}[b]{0.14\textwidth}
        \centering
        \includegraphics[width=\textwidth]{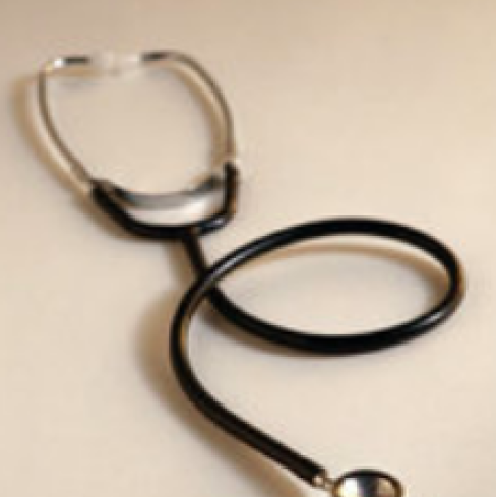}
        \caption{hang/wear/grab}
    \end{subfigure}
    \vskip\baselineskip
    \begin{subfigure}[b]{0.14\textwidth}
        \centering
        \includegraphics[width=\textwidth]{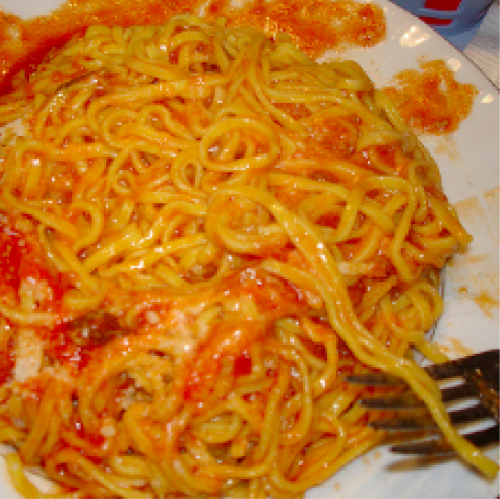}
        \caption{eat/taste} 
    \end{subfigure}
    \hspace{0.3cm}
    \begin{subfigure}[b]{0.14\textwidth}
        \centering
        \includegraphics[width=\textwidth]{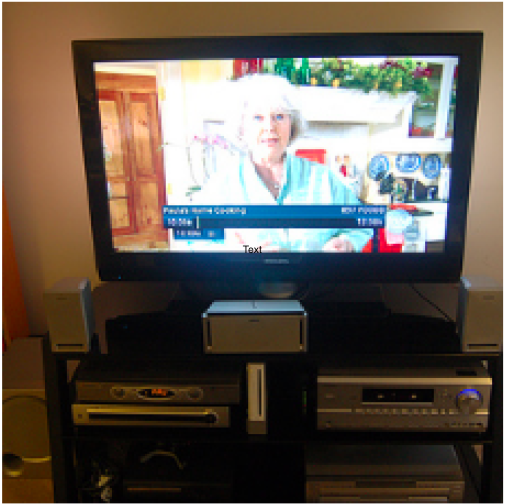}
        \caption{switch on/off}  
    \end{subfigure}
    \vskip\baselineskip
    \begin{subfigure}[b]{0.14\textwidth}
        \centering
        \includegraphics[width=\textwidth]{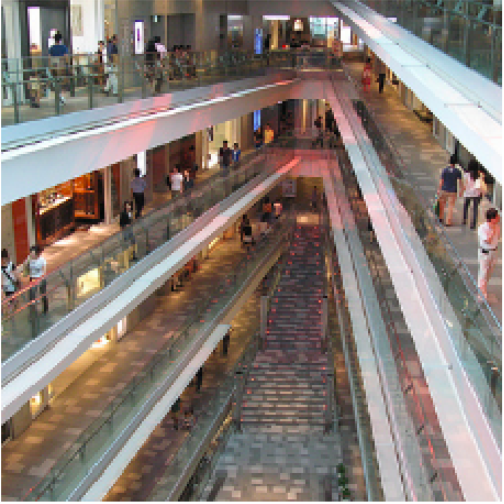}
        \captionsetup{format=hang}
        \caption{P: grab/drive\\
                 GT: walk/exit}
    \end{subfigure}%
    \hfill
    \begin{subfigure}[b]{0.14\textwidth}
        \centering
        \includegraphics[width=\textwidth]{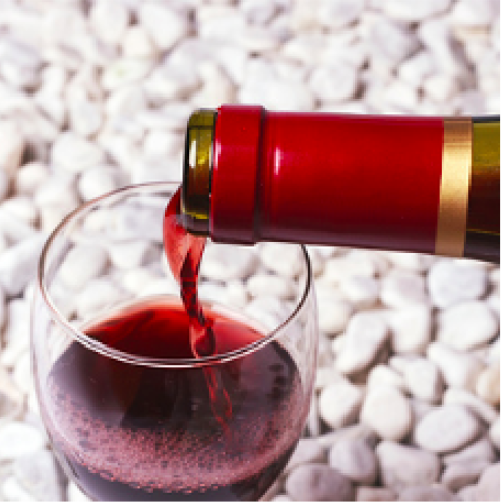}
        \captionsetup{format=hang}
        \caption{P: empty/fill \\ GT: taste}
    \end{subfigure}%
    \hfill
    \begin{subfigure}[b]{0.14\textwidth}
        \centering
        \includegraphics[width=\textwidth]{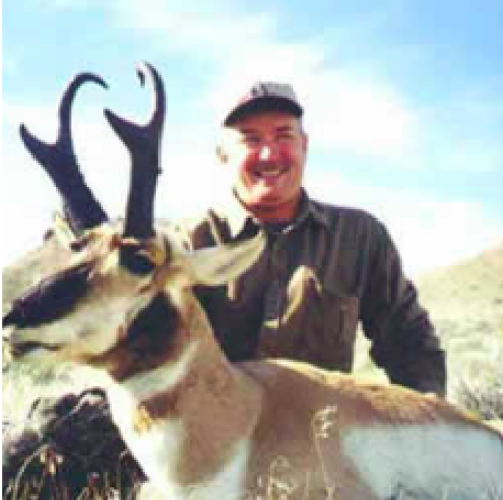}
        \captionsetup{format=hang}
        \caption{P: hunt/chase \\ GT: talk/meet}
    \end{subfigure}
\caption{Sample predictions. (a)-(d): Examples of images with correct affordance predictions (correct label below each image). (e)-(g): Examples of images with incorrect affordance predictions (P: indicates the CNN prediction, while GT: indicates the ground truth based on human labeling.}
\label{fig:sample_prediction}
\end{figure}

Figures~\ref{fig:sample_prediction}(a)--(d) demonstrate images where the network was able to predict correctly. However, the presence of distinct features can mislead the network. For example, in Figure~\ref{fig:sample_prediction}(e), where white bars stand out in the image, the network predicted ``grab'' and ``drive'', potentially mistaking the image as a bar or a road. On the other hand, human labelers, knowing that it is a image of a wall, provided labels such as ``walk'' and ``enter''. Since ImageNet contains natural scene images, multiple objects are likely to appear in one image, even though each image is assigned only one object label. Such images confuse both the labelers and the network, and therefore can lead to incorrect affordance recognition as shown in Figures~\ref{fig:sample_prediction}(f) and (g).

\section{Visualizing the Learned Representation Space}
\subsection{RDM across Layers}
\begin{figure}[!h]
    \centering
    \includegraphics[width=0.45\textwidth]{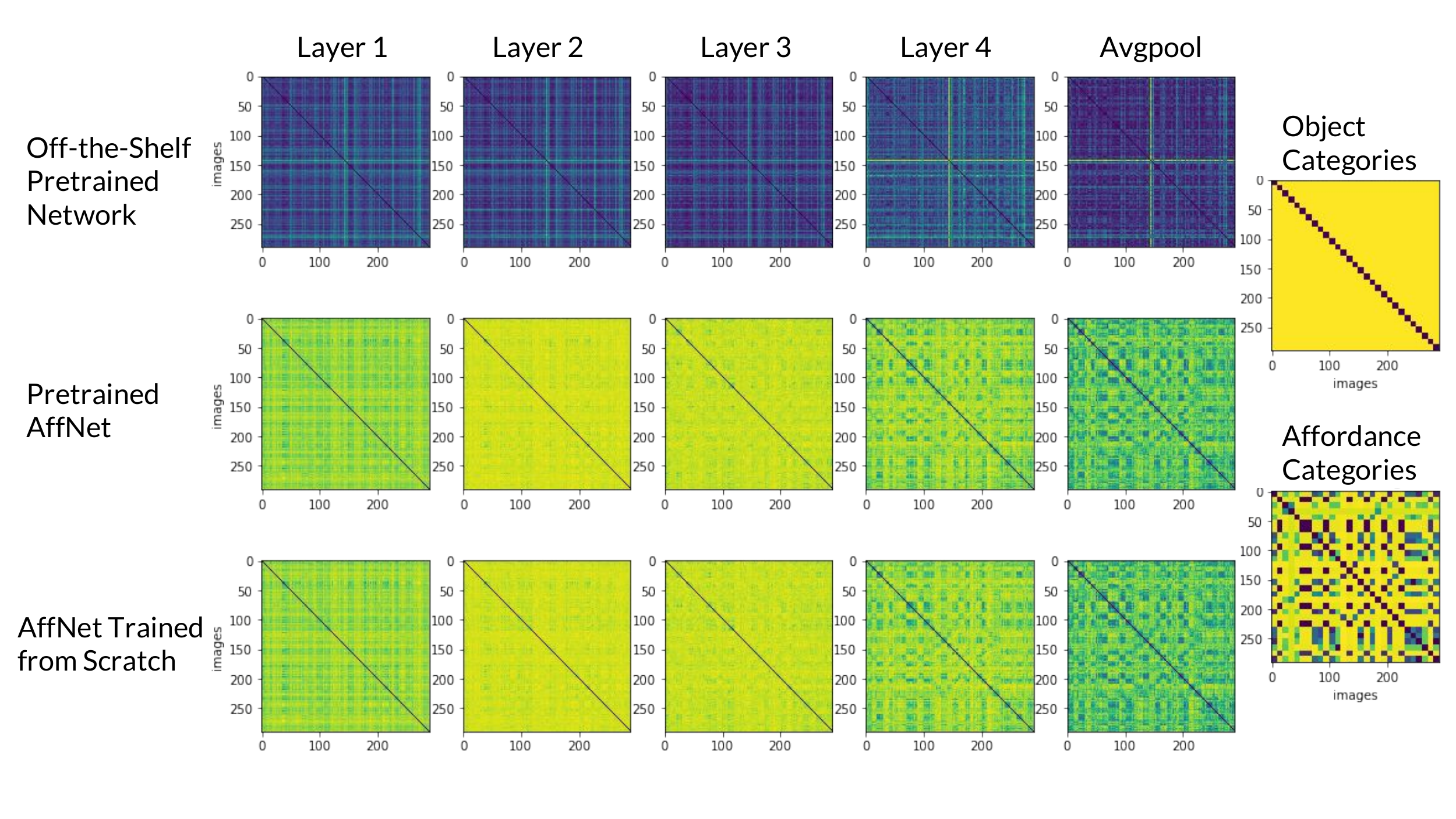}
    \caption{RDM matrix of layers from CNN from off-the-shelf pre-trained network, fine-tuned network and network trained from scratch for affordances.}
    \label{fig:rdm}
\end{figure}

To visualize the representations learned by the network, we randomly sampled 10 images from each of 30 objects classes, and extracted activations from the network layers. Pairwise correlation distance between network activation across layers was computed for each pair of images, and is shown in Figure~\ref{fig:rdm}. Pairwise distance between affordance labels is shown in the bottom-right matrix. This matrix denotes the ground truth distance in affordance space. Similar patterns begin to emerge in Layer 4 for both the fine-tuned network and the network trained from scratch. Critically, this pattern is not seen for the off-the-shelf network that was not trained on affordances. This demonstrates that our network learns representations that effectively separate different affordance categories.
\subsection{t-SNE}
\begin{figure}[!h]
    \centering
    \includegraphics[width=0.45\textwidth]{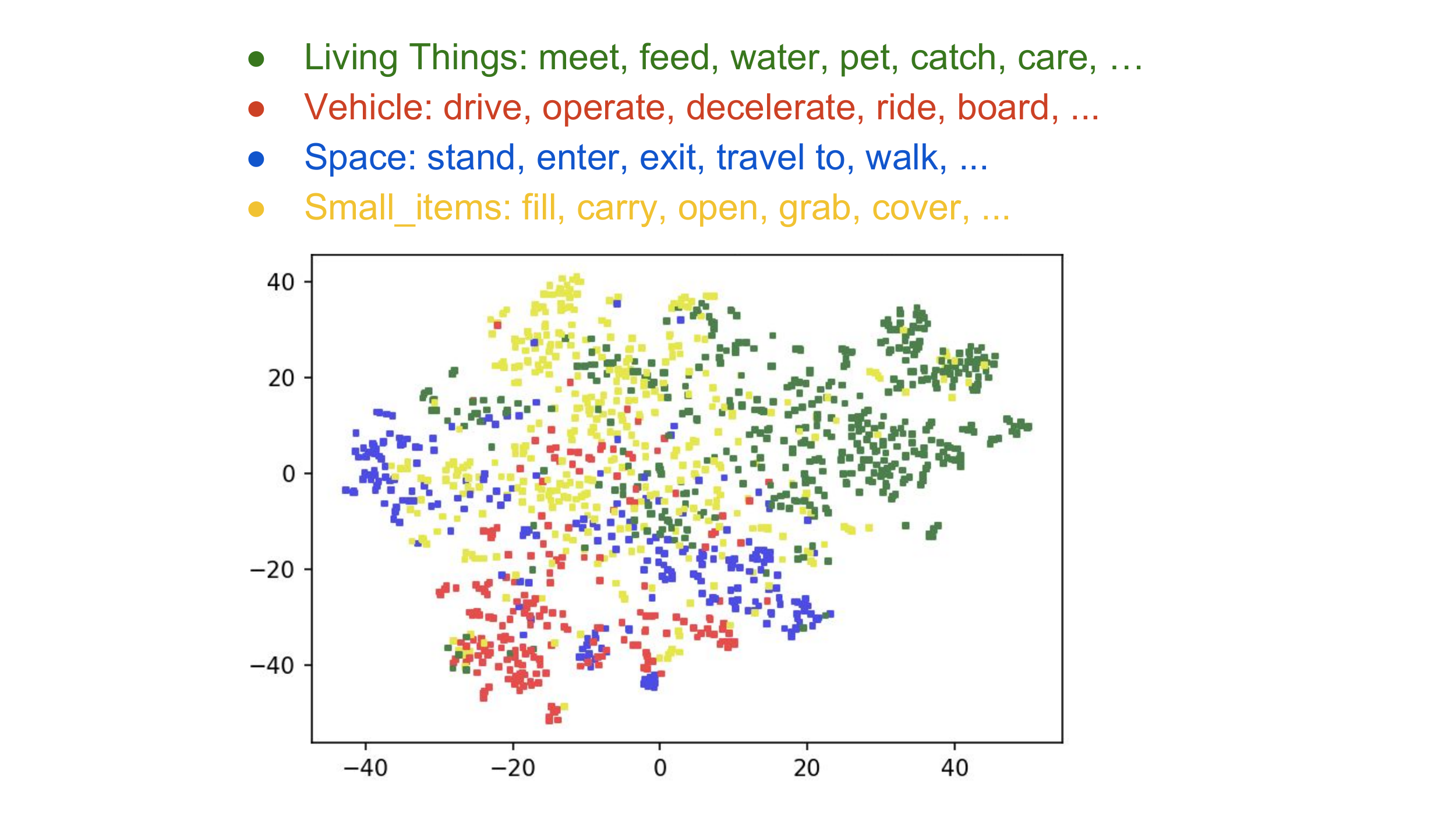}
    \caption{t-SNE visualization of the second to last layer in the CNN trained from scratch. Representations of images are coarsely split into four groups based on the distinct affordances of the images: living things (green), vehicles (red), physical spaces (blue) and small items (yellow).}
    \label{fig:tsne}
\end{figure}

Activations from the second to last layer in the network trained from scratch were visualized using t-SNE~\cite{maaten2008visualizing}, as shown in Figure~\ref{fig:tsne}. Images are coarsely split into four groups based on their distinct affordances: living things, vehicles, physical spaces and small items. In the 2D t-SNE visualization, the representation of living things (in green), vehicles (in red) and physical spaces (in blue) are visibly separable. Small items (in yellow), in contrast, span the entire space. The category of small items does not appear well separated, which is likely due to the visualization being limited to 2 dimensions.

\subsection{Unit Visualization}
\begin{figure}[!h]
    \centering
    \begin{subfigure}[b]{0.13\textwidth}
        \centering
        \includegraphics[width=\textwidth]{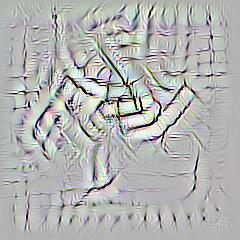}
        \caption{Ride}
    \end{subfigure}
    \hspace{0.2cm}
    \begin{subfigure}[b]{0.13\textwidth}
        \centering
        \includegraphics[width=\textwidth]{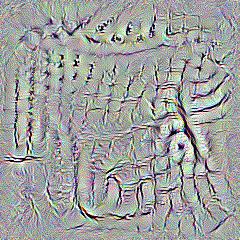}
        \caption{Wear}
    \end{subfigure}
    \hspace{0.2cm}
    \begin{subfigure}[b]{0.13\textwidth}
        \centering
        \includegraphics[width=\textwidth]{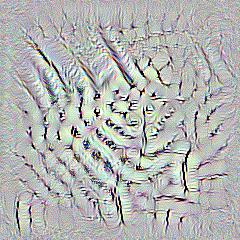}
        \caption{Watch}
    \end{subfigure}
    \vskip\baselineskip
    \begin{subfigure}[b]{0.13\textwidth}
        \centering
        \includegraphics[width=\textwidth]{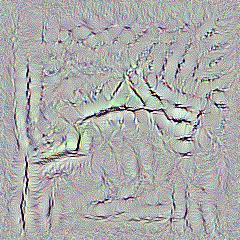}
        \caption{Climb}
    \end{subfigure}
    \hspace{0.2cm}
    \begin{subfigure}[b]{0.13\textwidth}
        \centering
        \includegraphics[width=\textwidth]{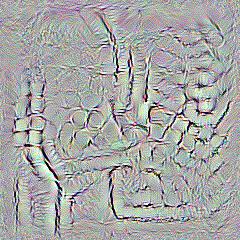}
        \caption{Fill}  
    \end{subfigure}
    \hspace{0.2cm}
    \begin{subfigure}[b]{0.13\textwidth}
        \centering
        \includegraphics[width=\textwidth]{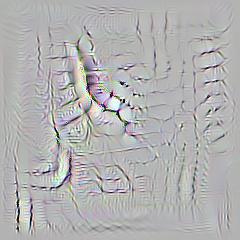}
        \caption{Sit}  
    \end{subfigure}
\caption{Visualization of 6 last layer units in the CNN.}
\label{fig:unit}
\end{figure}

We were able to visualize the output layer units of the CNN by optimizing in pixel space to determine which images maximally activated a specific unit. Figure~\ref{fig:unit} shows such visualization of 6 units from the output layer. The ``ride'' unit, for example, shows human- and horse-like structures; the ``wear'' unit shows a coarse clothing pattern and details of common textures often associated with clothing. Similarly, units ``climb'', ``sit'', and ``fill'' show stairs-like, chair-like, and container-like structures respectively. Interestingly,  the ``watch'' unit shows preference for dense textures in the center of the image space, which may correlate with image characteristics from objects that are related to watching (e.g., TV). It should be noted that unit visualization is very limited for capturing the learned intermediate features. Interpreting features in a limited 2D space is inherently biased and subjective. 

\section{Discussion}
We successfully trained a CNN to predict affordances from images, as a means for learning the underlying dimensionality of object affordances. The intermediate features in the CNN constitute an underlying compositional structure for the representation of affordances. 

\section{Future Work}
To ensure the objectivity of the affordance labeling, affordance labels for images -- as opposed to just object categories labels -- are being collected currently using Amazon Mechanical Turk. This dataset will be made publicly available after verification. 

With a CNN trained for affordance recognition, weights from the intermediate layers can be extracted and used to featurize each image. A model can then be trained to predict the BOLD responses to each image. Correlations between the predicted responses and the true responses can be used to measure model performance. If a linear model is built to perform this task, the model weights could then be used as a proxy to localize where information about affordances is represented in the human brain. 

Finally, affordance categories can be split into two large groups: semantically relevant ones, such as ``eat'', which requires past experience with the objects in question; and non-semantically relevant ones, such as ``sit'', which may be inferred directly from the shapes of the objects. If semantic affordances are being processed in the brain, top-down information about the objects is potentially necessary in order to inform an observer about affordable actions, while the non-semantic ones would not require top-down information. Given such differences we may be able to differentiate between top-down and bottom-up visual processing in the human brain using our model; in particular, by distinguishing the different brain regions that are engaged in either or both of these two processes.

\section{Acknowledgments}
This project is funded by MH R90DA023426-12 Interdisciplinary Training in Computational Neuroscience (NIH/NIDA)
\bibliographystyle{apacite}

\setlength{\bibleftmargin}{.125in}
\setlength{\bibindent}{-\bibleftmargin}

\bibliography{reference}

\end{document}